\RequirePackage{fix-cm}
\documentclass[smallextended]{svjour3}       
\smartqed  
\pdfoutput=1 

\usepackage{subcaption}
\usepackage{booktabs} 
\usepackage{graphicx}
\graphicspath{{./}} 
\usepackage{times}
\usepackage{textcomp}
\usepackage{mathptmx}      
\begin{document}

\title{Measurement of azimuthal dependent muon flux by 2\,m\,$\times$\,2\,m RPC stack at IICHEP-Madurai}
\titlerunning{Azimuthal dependence of muon flux at IICHEP-Madurai}        

\author{S. Pethuraj \and G. Majumder \and V.M. Datar \and N.K. Mondal \and K.C. Ravindran \and B. Satyanarayana}

\institute{ S. Pethuraj \at
  Homi Bhabha National Institute,
  Mumbai-400094\\
  \email{s.pethuraj@tifr.res.in}           
  \and
  S. Pethuraj \and G. Majumder \and V.M. Datar \and K.C. Ravindran \and B. Satyanarayana \at
  Tata Institute of Fundamental Research,
  Mumbai-400005 \\
  \email{gobinda@tifr.res.in, vivek.datar@tifr.res.in, kcravi@tifr.res.in, bsn@tifr.res.in}
  \and
  N.K. Mondal \at
  Saha Institute of Nuclear Physics,
  Kolkata-700064\\
  \email{nkm@tifr.res.in}
}

\date{Received: date / Accepted: date}

\maketitle
\begin{abstract}
 The proposed 50 \,kton\, INO-ICAL experiment is an upcoming underground high energy physics experiment planned to be commissioned at Bodi hills near Theni, India ($9^{\circ}57'N$, $77^{\circ}16'E$) to study various properties of neutrino oscillations using atmospheric neutrinos produced by extensive air shower phenomenon. The resistive plate chamber has been chosen as the active detector element for the proposed INO-ICAL. An experimental setup consisting a stack of 12 layers of glass resistive plate chambers each with a size of $\sim$2\,m$\times$2\,m has been built at IICHEP, Madurai to study the performance and long-term stability of the resistive plate chambers(RPCs) commercially produced in large quantities by the Indian industries as well as its electronics for the front-end and subsequent signal processing. In this study, the azimuthal dependence of muon flux at various zenith angles at Madurai (9$^{\circ}$56'N, 78$^{\circ}$00'E and at an altitude of 160\,m above mean sea level) has been presented along with the comparison of Monte Carlo from CORSIKA and HONDA predictions.
\end{abstract}

\section{Introduction}
\label{sec:intro}

The cosmic rays coming from outside of the solar system mainly consists of primary protons and a small fraction of helium and heavier nuclei. The primary cosmic rays interact with the earth's atmosphere and produce a secondary shower of particles. The study of cosmic rays leads to the understanding of the universe. The primary cosmic rays coming more or less isotropically to the earth are modulated by the earth's magnetic field which causes asymmetry in intensity of their arrival direction. The primary particles having momentum more than the cut-off rigidity will reach the earth's atmosphere. The observed asymmetry in the cosmic rays due to the earth's magnetic field is called ``east-west'' asymmetry. The secondaries produced in the primary interaction have also been known to follow the asymmetry in the arrival direction. The east-west asymmetry of cosmic ray muons is mainly dependent on the momentum cutoff, latitude and altitude. This effect in cosmic ray muons has been studied by many experiments in the world\cite{dsrm},\cite{hanoi}, \cite{thomson} and \cite{shuhei}. The primary interaction of the cosmic ray mainly produces pions ($\pi^0$, $\pi^+$ and $\pi^-$). The neutral pions mostly decay to $\gamma\gamma$, whereas the charge pions $\pi^{+}$ ($\pi^{-}$) dominantly decay to $\mu^{+}$ ($\mu^{-}$) + $\nu_{\mu}$ ($\overline \nu_{\mu}$). The muons are the most dominant charged particles which are observed at sea level from the cosmic ray showers.

The INO-ICAL is going to be an underground experiment to study neutrino oscillation parameters \cite{inowhite} to resolve the neutrino mass hierarchy problem using atmospheric neutrinos. For neutrino oscillation studies it is vital to understand the ratio of ($\nu_{\mu}$ + $\overline\nu_{\mu}$)/($\nu_{e}$ + $\overline\nu_{e}$) and the angular dependence of the neutrino flux at the experimental site. The main sources of uncertainty to the flux of neutrinos on ground comes from the interaction models. The muon flux on ground offers a possibility to reduce such uncertainties. The pion decay is the main source of neutrinos below 1\,TeV energies. The experimentally recorded muon flux at various locations on the earth has been used to tune the hadronic interaction models to reduce the uncertainty in the production of pions in the primary interaction. The calibration procedure for the interaction models and the detailed calculation of the neutrino flux has been explained in Honda et al. \cite{Honda1}. This is the main motivation to study the azimuthal dependence of muon flux at the present experimental site which is just 100\,km east from the INO-site and near to the geomagnetic equator.

A 12-layer prototype of the INO-ICAL detector has been constructed using RPCs of about 2\,m\,$\times$\,2\,m in dimensions \cite{santonico} at the IICHEP Transit Campus at Madurai, Tamilnadu in the southern part of India. The performance and long-term stability of the RPCs have been extensively studied using the cosmic ray muons since the inception of the detector. The principal aim of this study is to observe the azimuthal dependence of cosmic ray muons at different zenith angles using this prototype RPC stack and compare it with the various phenomenological models to tune parameters of those models and consequently better estimation of neutrino flux. In this paper, the experimental setup is explained in section \ref{sec:setup}. A preliminary analysis of event data is described in section \ref{sec:analysis} followed by the GEANT4 based MC technique, which is used for detector simulation in section \ref{sec:mcgen}. The estimation of muon flux and its systematic uncertainties due to various parameters has been discussed in section \ref{sec:aziflux} and section \ref{sec:syst} respectively. Lastly, the observed results along with its comparison with the CORSIKA and HONDA flux models are presented in section \ref{sec:compar}.

\section{The prototype detector setup}
\label{sec:setup}
\begin{figure}[h]
  \begin{center}
    \includegraphics[width=0.7\linewidth]{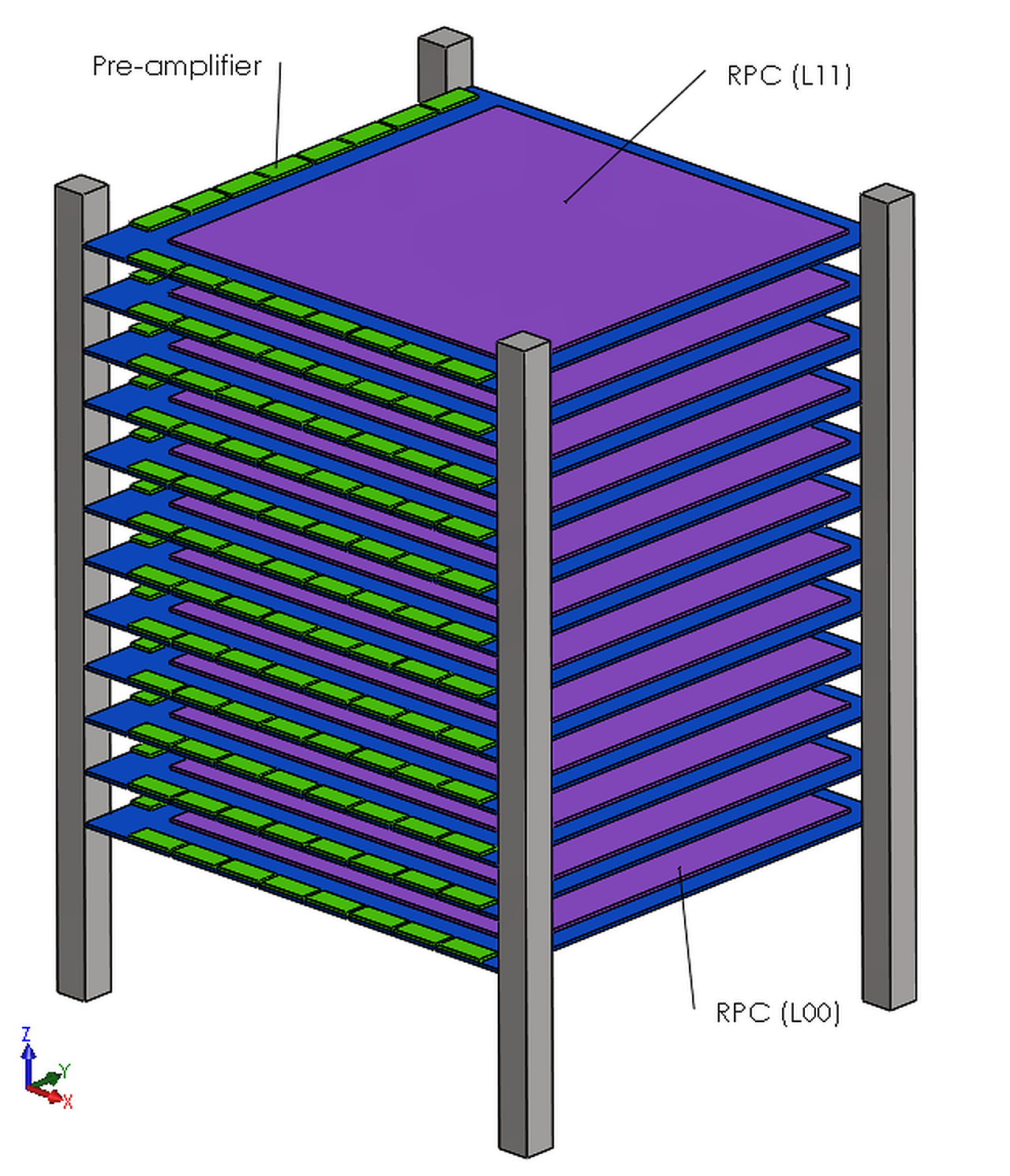} 
    \caption{12-layer detector stack of 2\,m\,$\times$\,2\,m RPCs.}
    \label{fig:detstack}
  \end{center}
\end{figure}

\begin{figure}[h]
  \includegraphics[width=1.0\linewidth]{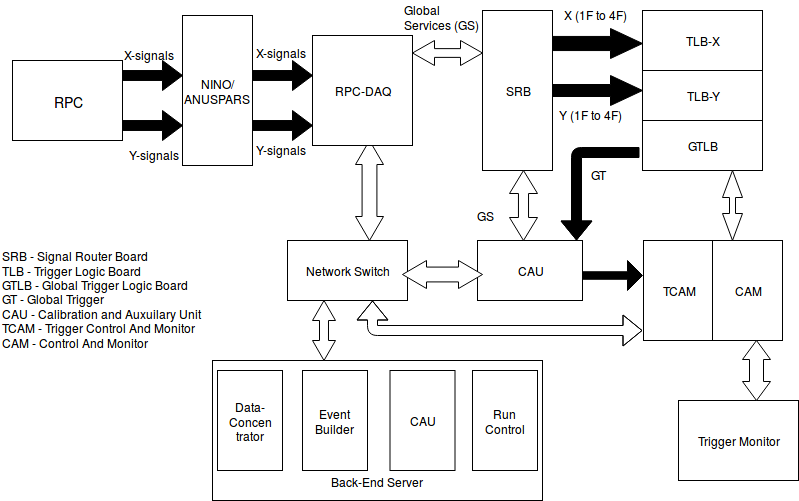} 
  \caption{Signal flow from RPC to Back-End.}
  \label{fig:sigflow}
\end{figure}

\begin{figure}[h]
  \includegraphics[width=1.0\linewidth]{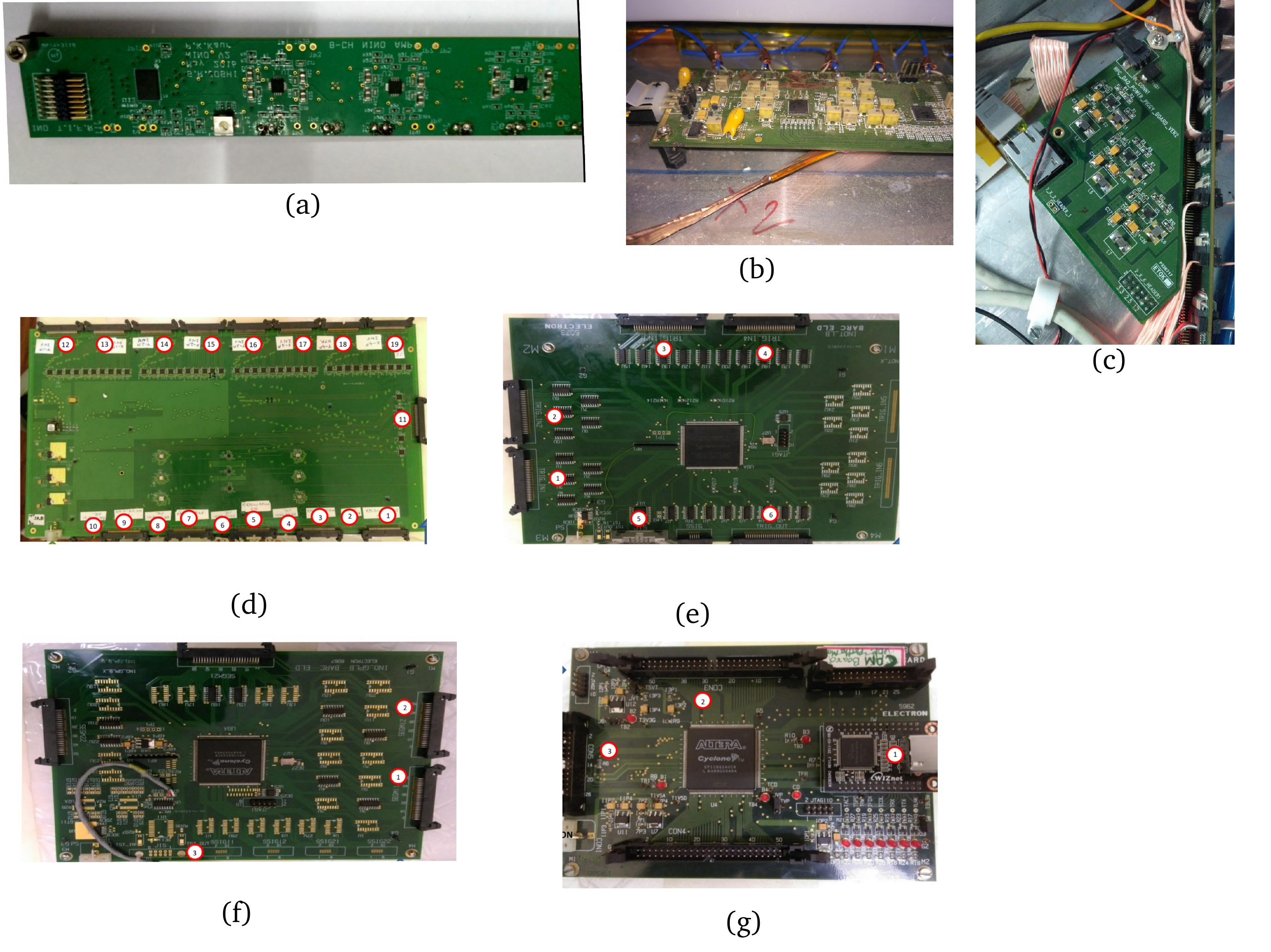} 
  \caption{(a) NINO Front End board, (b) Anusparsh Front End board, (c) RPCDAQ Digital-Front End board, (d) Signal Router Board, (e) Trigger Logic Board (X and Y), (f) Global Trigger Logic Board, (g) Control And Trigger Monitor and Trigger Control And Monitor. }
  \label{fig:elecphotos}
\end{figure} 

\begin{figure}[h]
  \begin{subfigure}{0.5\textwidth}
    \includegraphics[width=1.0\linewidth]{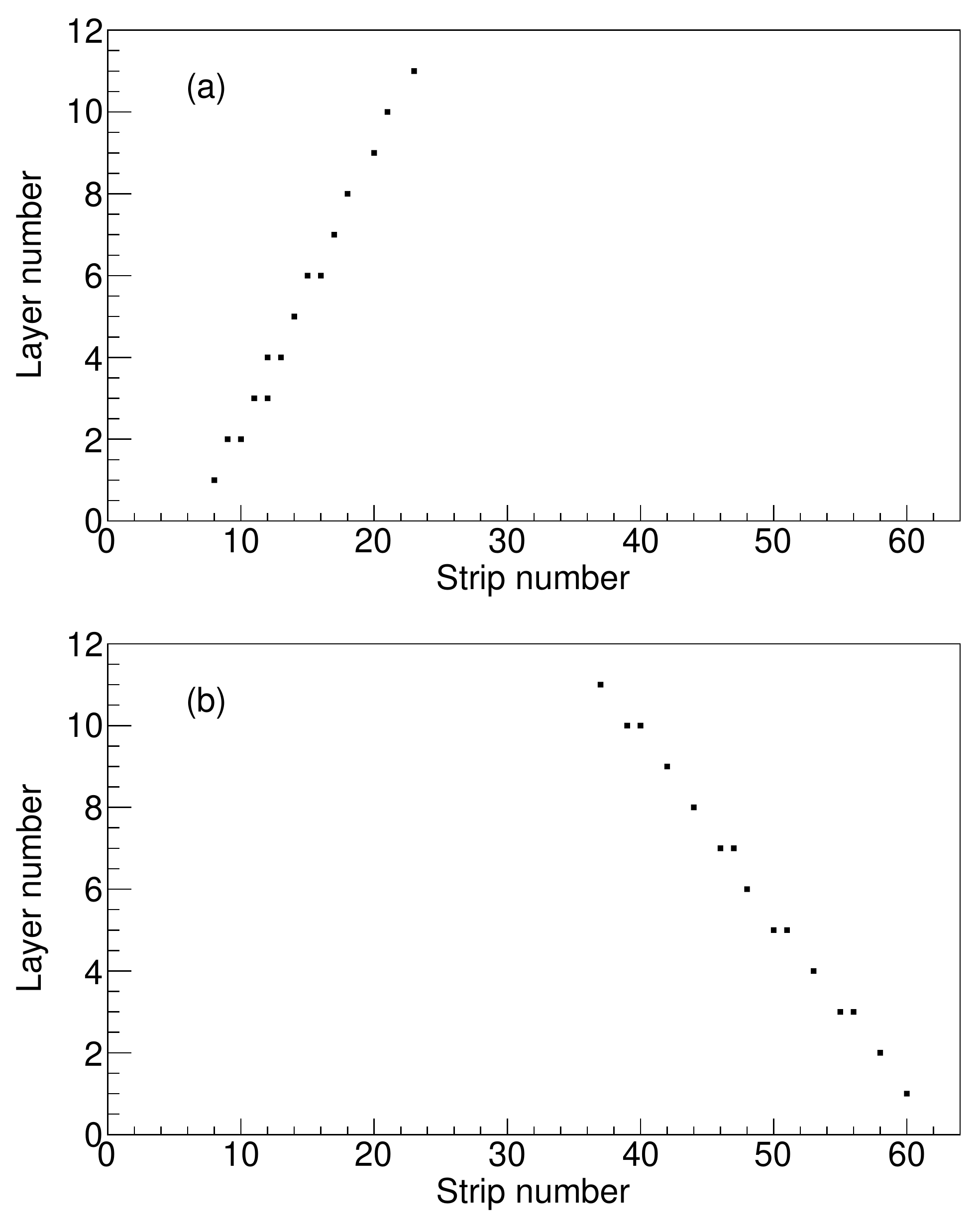} 
  \end{subfigure}
  \begin{subfigure}{0.5\textwidth}
    \includegraphics[width=1.0\linewidth]{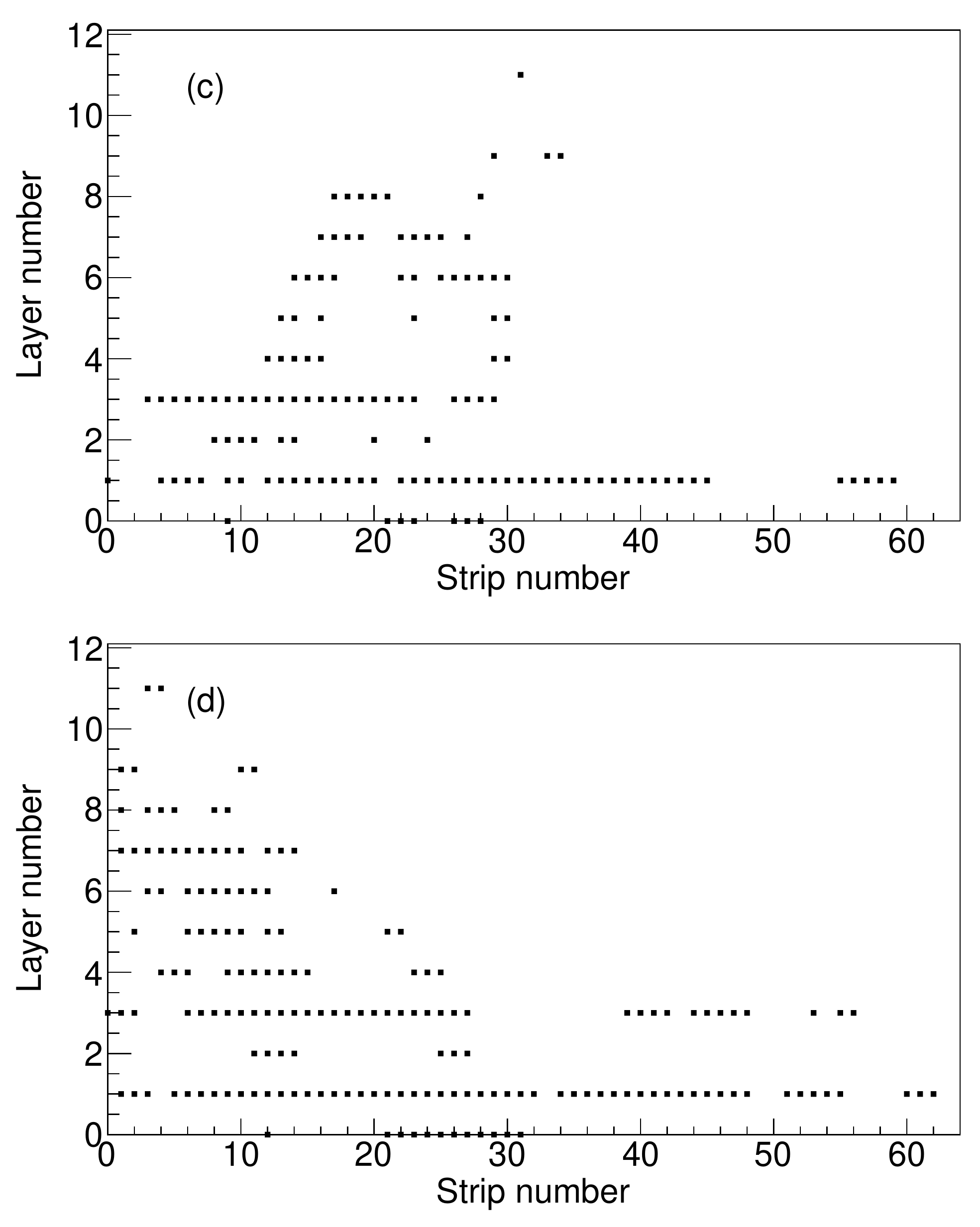}
  \end{subfigure}
  \caption{(a,b) Typical muon trajectory and (c,d) hadronic shower in RPC stack, (a,c) are XZ views and (b,d) are YZ view.}
  \label{fig:muonhadron}
\end{figure}

A graphical view of the 12 layer stack of RPCs is shown in figure \ref{fig:detstack}. The size of the RPCs used in the stack are 1.8\,m$\times$\,1.9\,m with an interlayer gap of $\sim$\,17\,cm. The Resistive Plate Chamber is a parallel plate chamber, built using two, 3\,mm glass plates which are separated by 2\,mm gap. The outer sides of the gas gap are coated by a thin layer of graphite paint in order to apply the high voltage contact. The RPC detector is operated in the avalanche mode by a continuous flow of gas mixture ($\mathrm{C_{2}H_{2}F_{4}}$ (95.2\,$\%$), iso-$\mathrm{C_{4}H_{10}}$ (4.5\,$\%$), $\mathrm{SF_{6}}$ (0.3\,$\%$)) with a differential bias voltage of $\pm$\,5\,kV. A charged particle passing through the RPC detector ionises the gas mixture, produces an avalanche and induces a fast signal on the external pickup strips. The RPC signals are readout by copper pickup panels, which are orthogonally (X- and Y- plane) placed on both sides of the chamber to record localized coordinates of the particle trajectory. The width of the pick-up strip is 28\,mm and the inter-strip gap between the strips is 2\,mm. The number of strips in X- and Y-plane are 60 and 63 respectively.

The induced signals from the pickup strip are amplified and discriminated by a charge sensitive NINO \cite{nino} Front-End Board (except layer 11, where Anusparsh \cite{anus1} voltage sensitive Front-End board is installed to study the performance of Anusparsh chips). The discriminated signals from NINO boards are passed to the FPGA-based Digital Front-End (RPCDAQ-board). The individual signals from every 8$^{th}$ strips (where the width of the discriminated signals are extended to 100\,ns) are ORed to get pre-trigger signals\footnote{ S$_{i}$ = CH$_{i}$ + CH$_{i+8}$ + CH$_{i+16}$ + CH$_{i+24}$ + CH$_{i+32}$ + CH$_{i+40}$ + CH$_{i+48}$ + CH$_{i+56}$ (CH$_{i+j}$ represents the (i+j)$^{th}$ strip ($i$ varies from 0 to 7) and ``+'' between two CH$_{i+j}$ denotes logical  ``OR'')} (S0 to S7). The four fold signals, namely the 1-fold\footnote{1F = S0+S1+S2+S3+S4+S5+S6+S7}, 2-fold\footnote{2F = S0$\cdot$S1+S1$\cdot$S2+S2$\cdot$S3+S3$\cdot$S4+S4$\cdot$S5+S5$\cdot$S6+S6$\cdot$S7}, 3-fold\footnote{3F = S0$\cdot$S1$\cdot$S2+S1$\cdot$S2$\cdot$S3+S2$\cdot$S3$\cdot$S4+S3$\cdot$S4$\cdot$S5+S4$\cdot$S5$\cdot$S6+S5$\cdot$S6$\cdot$S7} and 4-fold\footnote{4F = S0$\cdot$S1$\cdot$S2$\cdot$S3+S1$\cdot$S2$\cdot$S3$\cdot$S4+S2$\cdot$S3$\cdot$S4$\cdot$S5+S3$\cdot$S4$\cdot$S5$\cdot$S6+S4$\cdot$S5$\cdot$S6$\cdot$S7 (``+'' denotes logical  ``OR'' and  ``$\cdot$'' denotes logical  ``AND'')} are created by RPCDAQ which are passed to the Trigger system module in the back-end via Signal Router Board. The next level of trigger generation is performed in the various Trigger Logic Boards(TLBs). There are two TLBs for the X- and Y- view to observe the coincidence independently in these views. Four fold coincidence trigger is generated using 1-fold signal from four fixed trigger layer in X-plane. In parallel another four fold coincidence trigger is formed using 1-fold signals from four fixed trigger layers in Y-plane. The trigger signals of the X-Y planes are ORed in Global Trigger Logic Board (GTLB) to obatain the final trigger. The event data is recorded based on the arrival of the event trigger to the RPCDAQ from the back-end. As the trigger is generated in the back-end, there is a delay of $\sim$\,250\,ns between the time of passage of muon through the detector to the arrival of the trigger signal to the RPCDAQ. Hence, the pulse width of the event signal has to be increased to $\sim$\,300\,ns. In final ICAL experiment, the trigger signals will take much longer time (in few micro seconds) to reach RPCDAQ from the time of event occurred in the ICAL due to the larger cable length. The RPCDAQ is designed with provision to change the pulse width of the event signal. The RPCDAQ used in the present setup tested by increasing the pulse width up to 1\,$\mu$s. The event signals are latched in the RPCDAQs based on the trigger signal received from the back-end and are transferred to the data concentrator and Event Builder via Network Switch. The output of the event builder is stored in a local computer, which is then used for analysis. The flow of signals from the RPCs to the Back-End is shown in figure \ref{fig:sigflow}. The different electronics modules are shown in figure \ref{fig:elecphotos}. The detailed description of signal processing and Data Acquisition system (DAQ) can be found in \cite{elec1}, \cite{mandar1} and \cite{nagraj1}. The cosmic muons were recorded using the 1-Fold signals from layers 4, 5, 6 and 7 as a trigger. For this study, approximately 92 million events were recorded over a period of about $\sim$\,5\,days with an average trigger rate of 220\,Hz. Assuming the average energy loss of muons to be $\sim$\,2\,MeVg$^{-1}$cm$^{2}$, the minimum momentum cut off to reach the bottom trigger layer in the vertical direction is estimated to be $\sim$\,100\,MeV. This minimum momentum cutoff is mainly contributed by the roof of the building and the various supporting structures of the RPC stack. 

\section{Analysis of Experimental data}
\label{sec:analysis}
\begin{figure}
  \includegraphics[width = 1.\linewidth]{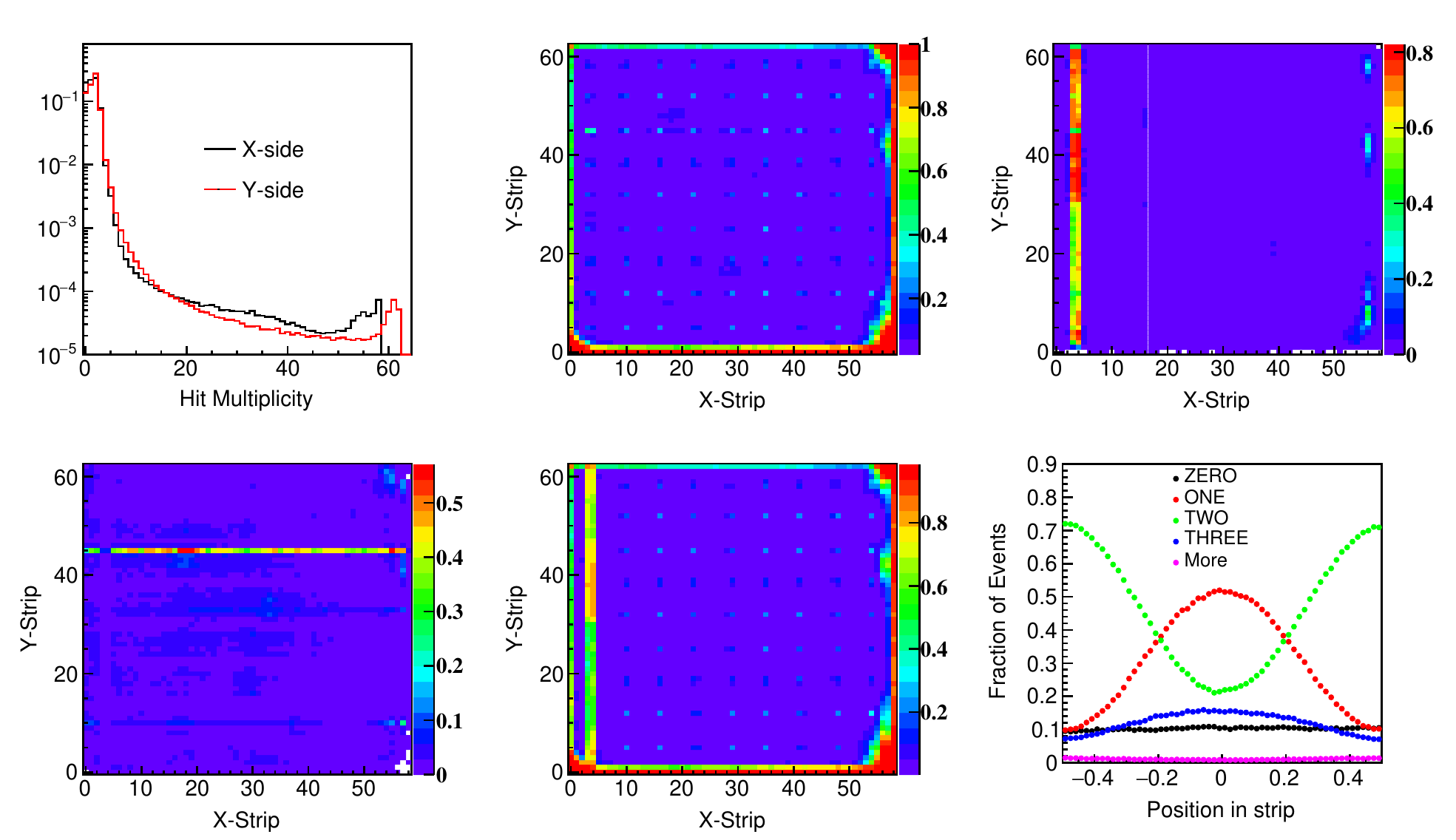}
  \caption{ (a) Strip-hit multiplicity in X-plane and Y-plane in Layer 2., 
    (b) Pixel-wise correlated inefficiencies in Layer 2,
    (c) Pixel-wise uncorrelated inefficiencies in Layer 2 (X-plane), 
    (d) Pixel-wise uncorrelated inefficiencies in Layer 2 (Y-plane), 
    (e) Pixel-wise trigger inefficiencies in Layer 2 (X-plane) and 
    (f) strip multiplicity based on muon hit position in a strip in Layer 2 (Y-plane). }
  \label{fig:image2}    
\end{figure}  

The event data consist of two information (i) cosmic ray muon hits\footnote{Hit is the induced strip signal above the set threshold after preamplifier.} per layer and (ii) the corresponding time of arrival. The strip hit information is stored as 1 bit per strip. For timing information, the signal from every 8$^{th}$ strips are combined and recorded by the TDC. The typical hit patterns observed from muon and hadron shower are shown in figure \ref{fig:muonhadron}. The average strip multiplicity for a signal induced by passage of muon is about 1.3 strip. However, there are a small fraction of events observed with large multiplicities which are mainly due to streamer signals, hadron shower or correlated electronic noise. A typical strip multiplicity plot is shown in figure \ref{fig:image2}(a). The zero hit in strip multiplicity is due to the inefficiency of the detector. For strip multiplicity of one or two, the observed position resolution is $\sim$\,7\,mm. For strip multiplicity of more than 3, the position resolution is of the order of the strip width.
\begin{figure}
  \center
  \includegraphics[width = 0.45\linewidth]{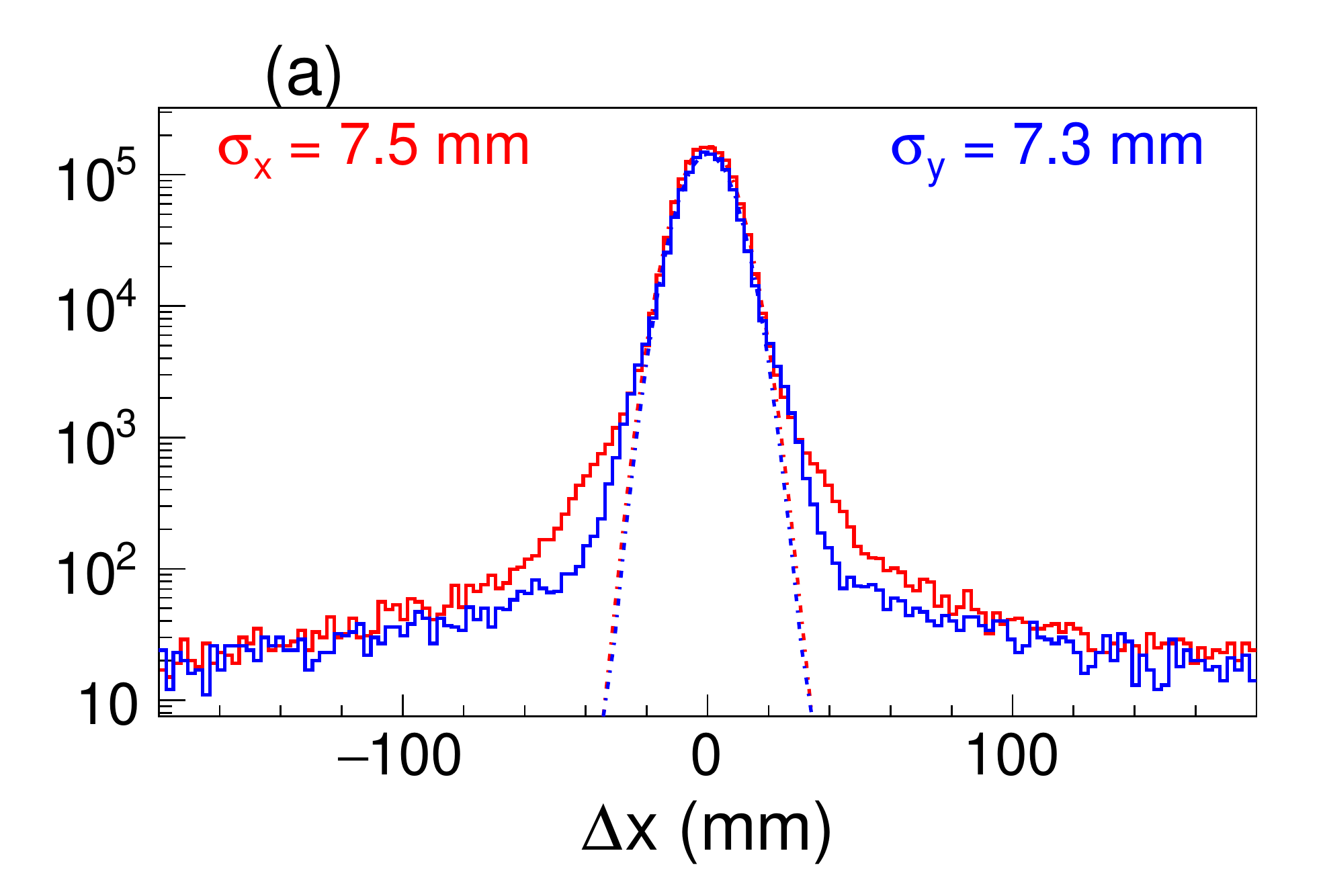}
  \includegraphics[width = 0.45\linewidth]{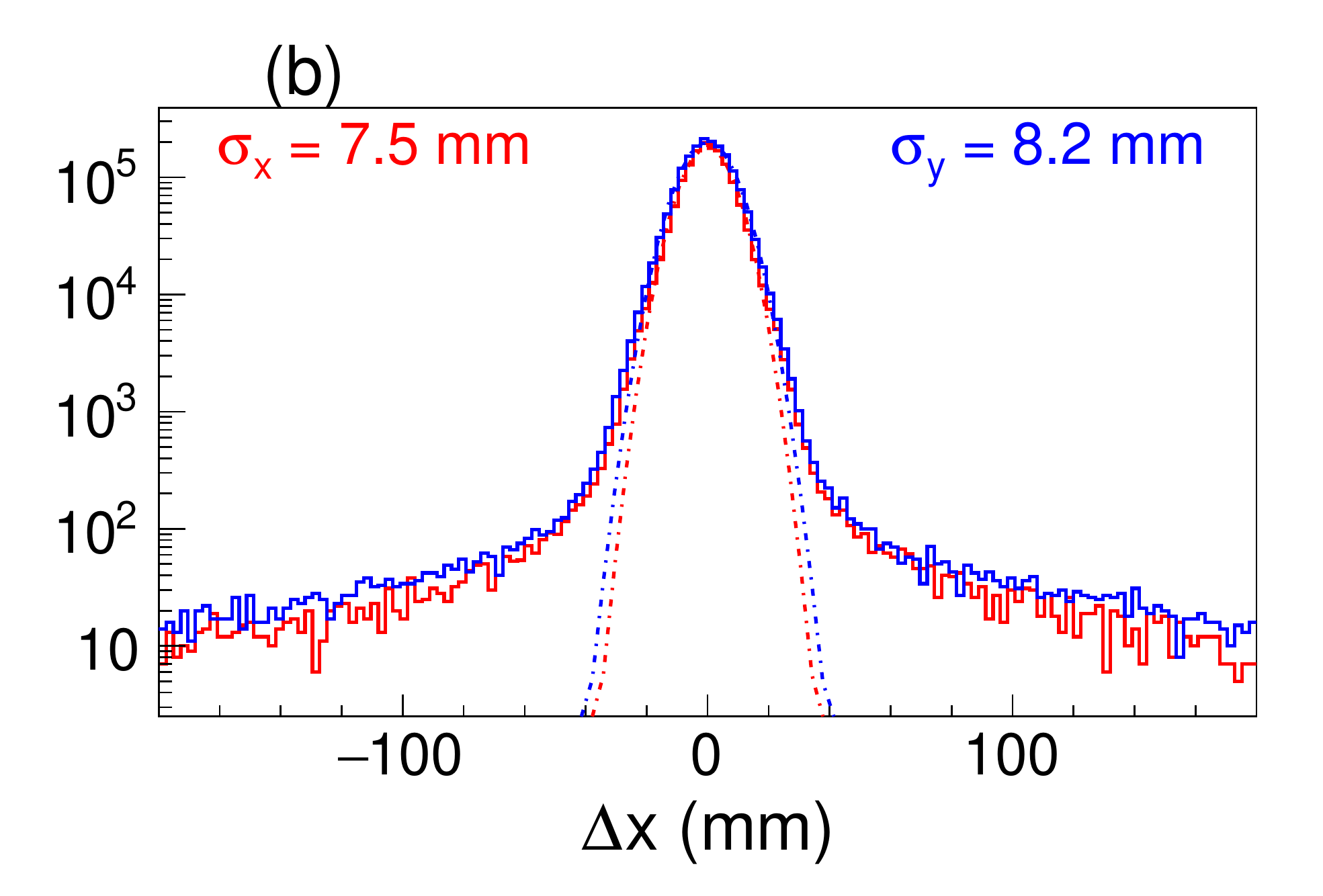} \\
  \includegraphics[width = 0.45\linewidth]{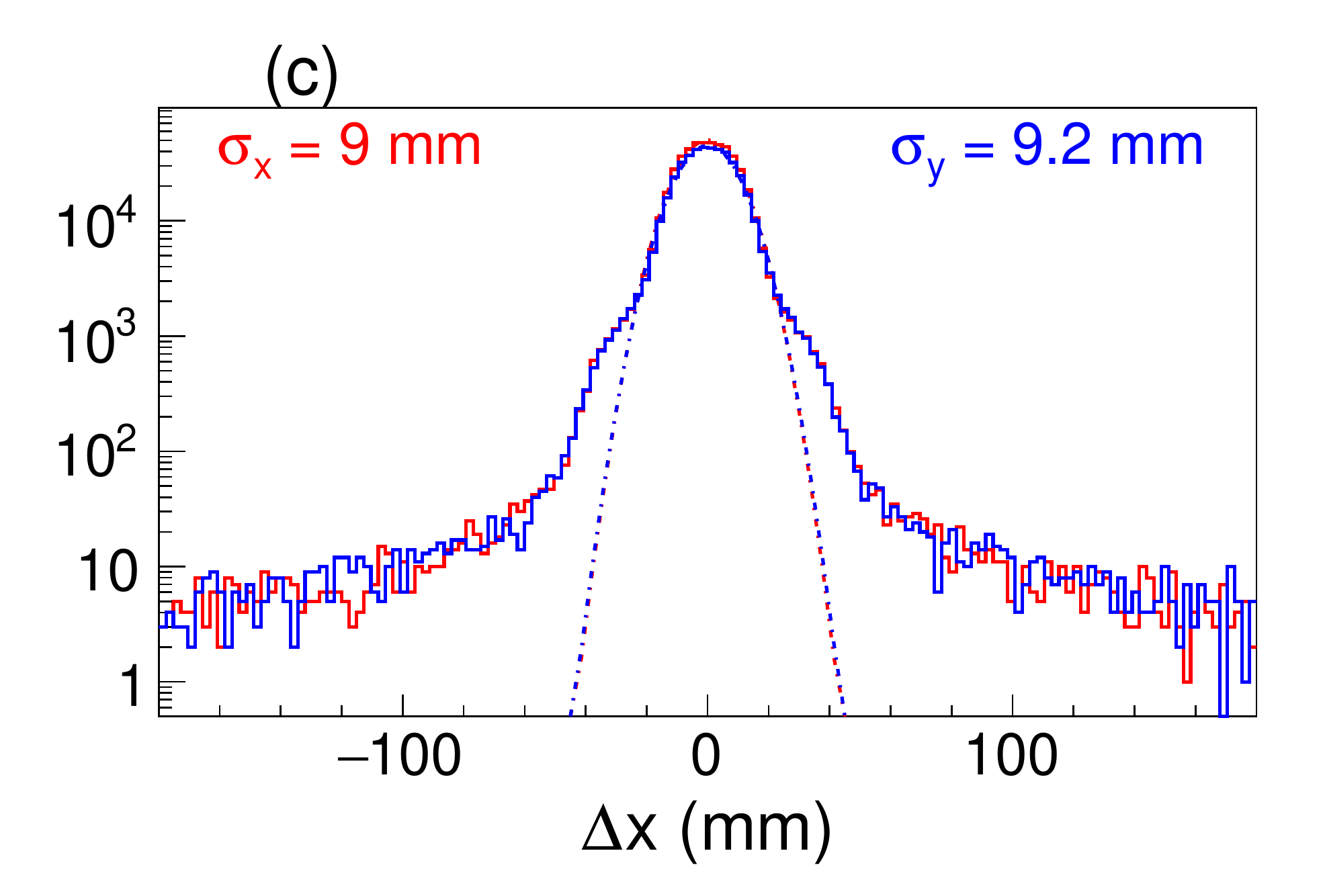}
  \includegraphics[width = 0.45\linewidth]{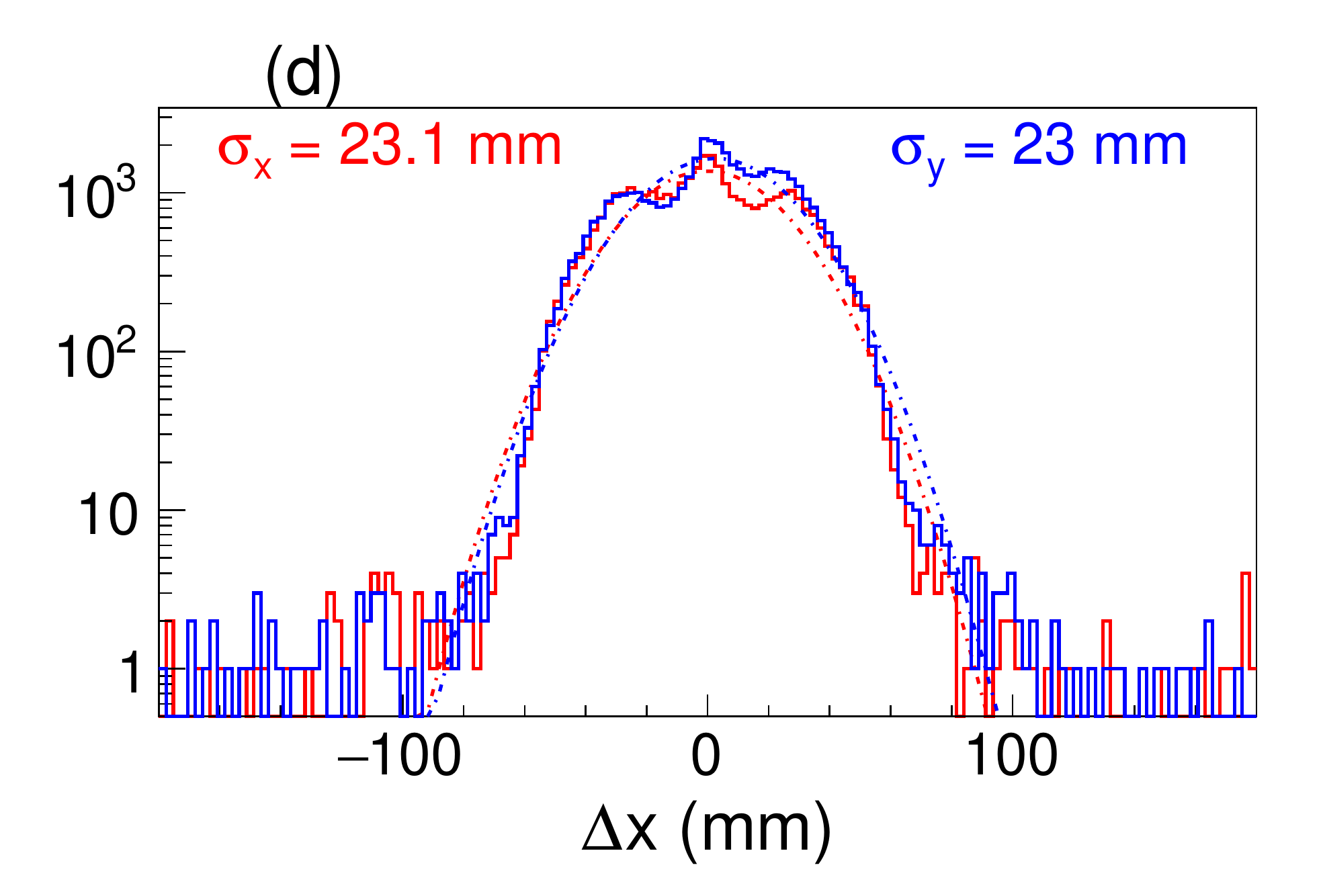}
  \caption{(a), (b), (c) and (d) are position residues of X- and Y-plane of RPC for strip multiplicity one, two, three and four respectively.}
  \label{fig:posreso}
\end{figure}
Due to poorer position resolution at the large strip multiplicities, the layer with atmost three consecutive strip hits in X- or Y- plane is only used in the current study. The selected hit position (arithmetic mean of all position) from all layers are fitted using the straight line in XZ- and YZ-plane using the equation \ref{eqn:strght},

\begin{equation}
  \label{eqn:strght}
  x(/y) = \alpha \times z + \beta
\end{equation}
where $x$ or $y$ is the hit location from the X- or Y-plane respectively for Z$^{th}$ layer, \textbf{$\alpha$} is the slope which is $\mathrm{tan\theta cos\phi}$ ($\mathrm{tan\theta sin\phi}$ ) for XZ (YZ) plane and \textbf{$\beta$} is the intercept. The precise position of the muon track in all the RPC layers are computed using the above-mentioned fit parameters. During the commissioning of the detector, the RPCs were manually placed in the stack and were carefully aligned to have minimum deviation between different layers. But there is still some misalignment in the relative detector position of different layers. This small relative shift needs to be corrected to get accurate fit parameters. An iterative method has been developed to calculate these relative shifts using the muon data acquired by the stack.

In this iterative method, the layer, for which the alignment value to be calculated in the iteration, is excluded from the fit. The extrapolated hit position in the layer under study is computed using the fit parameters of the fit of position hits from the other layers and the difference between the measured and extrapolated hit position is calculated. The relative shift in the layer is observed from the mean position of the distribution of these differences for the entire dataset which is expected to be gaussian and peak around zero. The detector misalignment is observed by any shift of the peak position from zero. These corrections for all layers are done one-by-one by excluding that layer in fit. After four or five iterations, the accuracy of overall chamber alignment has been achieved to be better than 0.2\,mm. The detailed description of this technique along with the various selection criteria can be found in \cite{spal}. The distribution of X-residues and Y-residues are fitted with a gaussian function, the fitted gaussian $\sigma$ represents the position resolution of the RPCs. The position resolution at different strip multiplicity for  X- and Y-plane in RPC are presented in figure \ref{fig:posreso}.

The RPCs in the stack can be divided into a matrix of pixels each of size 3\,cm\,$\times$\,3\,cm by combination of all X and Y strips. The inefficiencies are observed for each of this pixel and represented as inefficiency map for each RPC. For the estimated point in the layer under study, the extrapolation error on the hit position is estimated as,
\begin{equation}
  \label{eqn:exterr}
\epsilon = \sqrt{\sigma_{\alpha}^{2}\,z^{2} + \sigma_{\beta}^{2} + 2\,z\,cov(\alpha,\beta)}
\end{equation}

Also, the deviation($\delta$) of the fit point from the midpoint of the strip is calculated. The trajectories where $|\delta|$ + $\epsilon$ is less than a strip pitch on both X- and Y-plane are selected for further analysis. For each pixel in the layer, the number of muons passed through it is estimated from the extrapolated position in that layer. From the observed data, the ratio of the number of events where no hit is observed in both X- and Y- strip of that pixel to the total number of expected events to pass through is considered as correlated inefficiency. From the observed data, if for a pixel either X- or Y- strip hit is missing, then the ratio of those events to the total number of expected muons is taken as uncorrelated inefficiency for the corresponding X- or Y- plane. The pixel-wise correlated inefficiencies of layer-2 are shown in figure \ref{fig:image2}(b). The correlated inefficiencies are used to study the properties of the gas gaps which corresponds to both X- and Y-plane, like the uniformity of the gas gap, button spacers and non-uniformity of the gain inside the gap. The uncorrelated inefficiencies for the X- and Y-plane in layer-2 are shown in figure \ref{fig:image2}(c) and \ref{fig:image2}(d) respectively. These are used to identify the plane dependent properties for a particular RPC detector like dead strips, low gain strips, non-uniformity in the surface resistivity of the graphite coating, etc,. To calculate the trigger inefficiency of a layer, the ratio of the number of events for a pixel where the hit is not found in any of the strip in that plane, to the total number of events is considered. The trigger inefficiency maps are used for the trigger layers (layer 4, 5, 6 and 7) in the MC simulation to accept the event. To avoid bias, the trigger inefficiencies of the top six layers are estimated by using the events triggered from the bottom four layers. Similarly, the trigger inefficiencies of the bottom six layers are estimated by using the events triggered form the top four layers. The trigger inefficiency of an RPC is shown in figure \ref{fig:image2}(e). The inefficient spots which are observed at an interval of 6 strips are due to button spacers placed inside the gas gaps to maintain the uniform thickness between the two glass electrodes.

For each RPCs, the strip multiplicity depends on the position of the trajectory within the strip and is shown in figure \ref{fig:image2}(f). From the figure, it can be observed that the probability of observing single strip multiplicity is maximum when muon passes through the centre of the strip. The probability for two strip multiplicity is more when the muon passes through the edges of the strips due to sharing of the charge by nearby strips. For the case of three strip multiplicity it can be seen that the maximum probability is at the centre of the strip similar to the case of single strip multiplicity. Although, the observation of strip multiplicity above three is less probable, it can be seen that it is more or less flat distribution. A similar analysis procedure has been applied in \cite{pethu1}, where the zenith angular dependence of cosmic ray muon and the vertical flux for the same location are discussed. 

\section{Monte-Carlo event generation}
\label{sec:mcgen}

The Monte-Carlo(MC) simulation for the present work has been performed in two steps. In the first step, the CORSIKA\cite{heck} software is used for the extensive air shower simulation and the secondary particle information are stored for the present experimental site. In the next step, the secondaries observed at the output of the CORSIKA are propagated through the detector simulation performed using the GEANT4 \cite{geant4} toolkit. In the CORSIKA simulation, twenty million primary protons and two million Helium are generated as primaries from the top of the atmosphere. The energy range of these primaries is from 10\,GeV to 1\,PeV with a spectral index of -2.7. The horizontal flat detector option is chosen in the CORSIKA simulation due to the flat geometry of the RPC stack. The earth's magnetic field at the experimental site ($B_{X}$ (horizontal) = 40.431\,$\mu$T and $B_{Z}$ (vertical) = 4.705\,$\mu$T) is also given as an input here. In the CORSIKA simulation, the models GHEISHA and SIBYLL (version 2.3c) are respectively selected as the low and high energy interaction models. The generated primary proton energy is accepted according to the rigidity cutoff in different ($\theta$,$\phi$) bins. The primaries with the energy above the rigidity cutoff are allowed to propagate. The position ($x$, $y$) and momentum ($P_{x}$, $P_{y}$ and $P_{z}$) component of the particles at the observation level (muons and charged hadrons with a minimum energy threshold of 100\,MeV) are extracted from the output of the CORSIKA simulation. The global position information of these particles are digitized in the square of 2\,m\,$\times$\,2m and stored as the local position (within the square) with respect to the centre of the square.

\begin{figure}
  \center
  \includegraphics[width=0.9\linewidth]{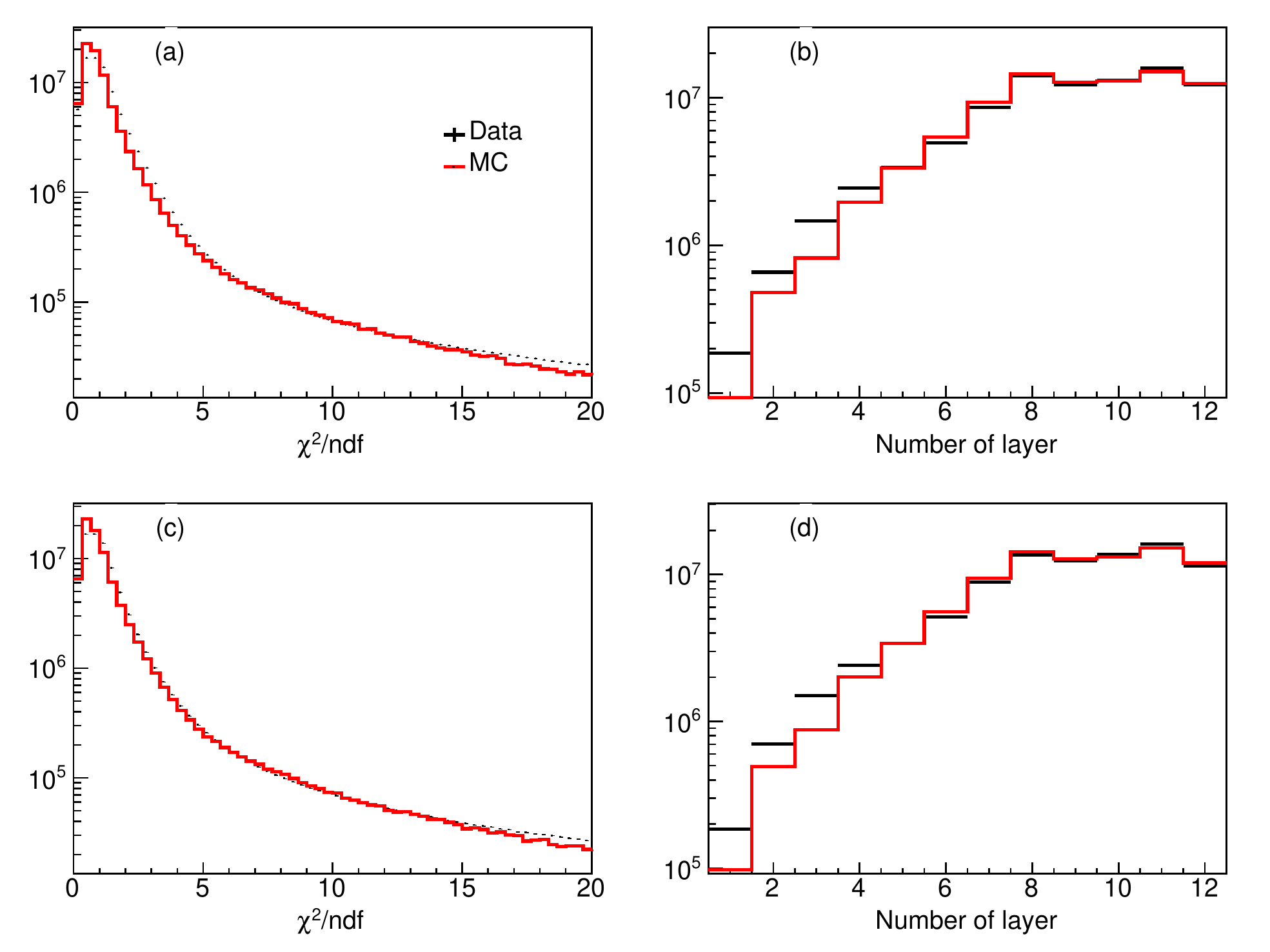} 
  \caption{(a,c) $\chi^{2}$/ndf for Data and MC in X- and Y-plane. (b,d) Number of layers in Data and MC for X- and Y-plane.} 
  \label{fig:chindf}
\end{figure}

To account for the passage of the particles through the materials of the detector, a detector simulation package has been developed using the GEANT4 simulation framework. The experimental hall along with nearby buildings and the 12 layer RPC stack inside the experimental hall are included in the geometry for the simulation. Here, the secondary particles are generated above the roof of the building. The various detector parameters like uncorrelated and correlated inefficiencies, trigger inefficiencies and strip multiplicity, which are estimated using the data sample, are incorporated during the digitization process of simulation. The steps followed in the MC event generation are, a position ($x,y$) and momentum component ($P_{x}, P_{y}$ and $P_{z}$) of the secondaries extracted from the output of CORSIKA is used to generate the particles on the topmost trigger layer (i.e., layer 7). The generated particle position is extrapolated to the bottom trigger layer to test the acceptance condition. The event generation vertex on top of the roof is calculated for the set of ($x,y,P_{x},P_{y},P_{z}$) and given as input to GEANT4. The simulation of the passage of a particle through the detector geometry is performed by the GEANT4. When the particle, passes through an RPC (sensitive detector volume), the GEANT4 provides the ($x,y,z$) coordinate and the exact time for that point. The digitization and incorporation of the detector parameters extracted from the data are explained in \cite{pethu1}. For the sake of completeness, the same steps are discussed further. The coordinates of the hit position in the global coordinates are translated into the strip information for the corresponding Z-plane. The pixel-wise correlated inefficiency map discussed in the previous section is used to incorporate the correlated inefficiencies in the simulated event. The position (hit position with respect to the strip centre) dependent strip multiplicity, as shown in figure \ref{fig:image2}(f) is used to generate the strip multiplicity depends on the hit position within a strip. The uncorrelated inefficiencies for X and Y strips are incorporated independently based on strip multiplicity using the inefficiency map discuss in the previous section. The trigger inefficiencies are incorporated only for the trigger layers (namely layers 4, 5, 6 and 7) in the X- or Y- plane to accept an event. In the experimental data, random noise hits due to electronics and multi-particle shower within the detector volume are also observed. These noise hits are also extracted from data and incorporated during the digitization process. The simulated events are analysed in the same procedure that is used for the experimental data. The comparison of $\chi^{2}$/ndf and the number of layers hit on both X- and Y- planes are shown in figure \ref{fig:chindf}.

\section{Estimation of muon flux at different ($\theta,\phi$) bins}
\label{sec:aziflux}
The number of muons events reconstructed which have $\chi^{2}$/ndf less than 8 and more than five layer muon hits used to estimate
the intensity of muons at various ($\theta,\phi$), the equation used to calculate the muon flux at different ($\theta,\phi$) bins given in equation \ref{eqn:equation2},
\begin{equation}
  \label{eqn:equation2}
  I_{\theta, \phi} = \frac{I_{data} }{\epsilon_{trig} \times \epsilon_{selec} \times \epsilon_{daq} \times T_{tot} \times \omega }  
\end{equation}

where, $I_{data}$ is the number of reconstructed muon at a ($\theta,\phi$) bin, $\epsilon_{trig}$ (number of events triggered out of accepted) is the trigger efficiency in that ($\theta,\phi$) bin, $\epsilon_{selec}$ (number of events reconstructed out of triggered) is the event selection efficiency in that ($\theta,\phi$) bin, $\epsilon_{daq}$ is the efficiency due to dead time in the data acquisition system, $T_{tot}$ is the total time taken to record the data (in seconds) including DAQ's dead time (0.5 ms/event) and $\omega$ is the accepted solid angle times the surface area, which is further defined as,

\begin{equation}
  \label{eqn:equation3}
  \omega = \frac{{\it AN}}{{\it N^{\prime}}}\int_{\theta_1}^{\theta_2} \textrm{cos}\theta \textrm{sin}\theta d\theta \times \int_{\phi_1}^{\phi_2} d\phi
\end{equation}

where, $A$ is the surface area of the RPC in the top triggered layer, $N$ is the number of events accepted at a ($\theta,\phi$) bin when the generated position on the top and bottom trigger layer is inside the detector, $N^{\prime}$ is the number of events generated on top trigger layer at ($\theta,\phi$) bin.

\section{Systematic studies}
\label{sec:syst}
\begin{figure}
  \center
  \includegraphics[width=1.0\linewidth]{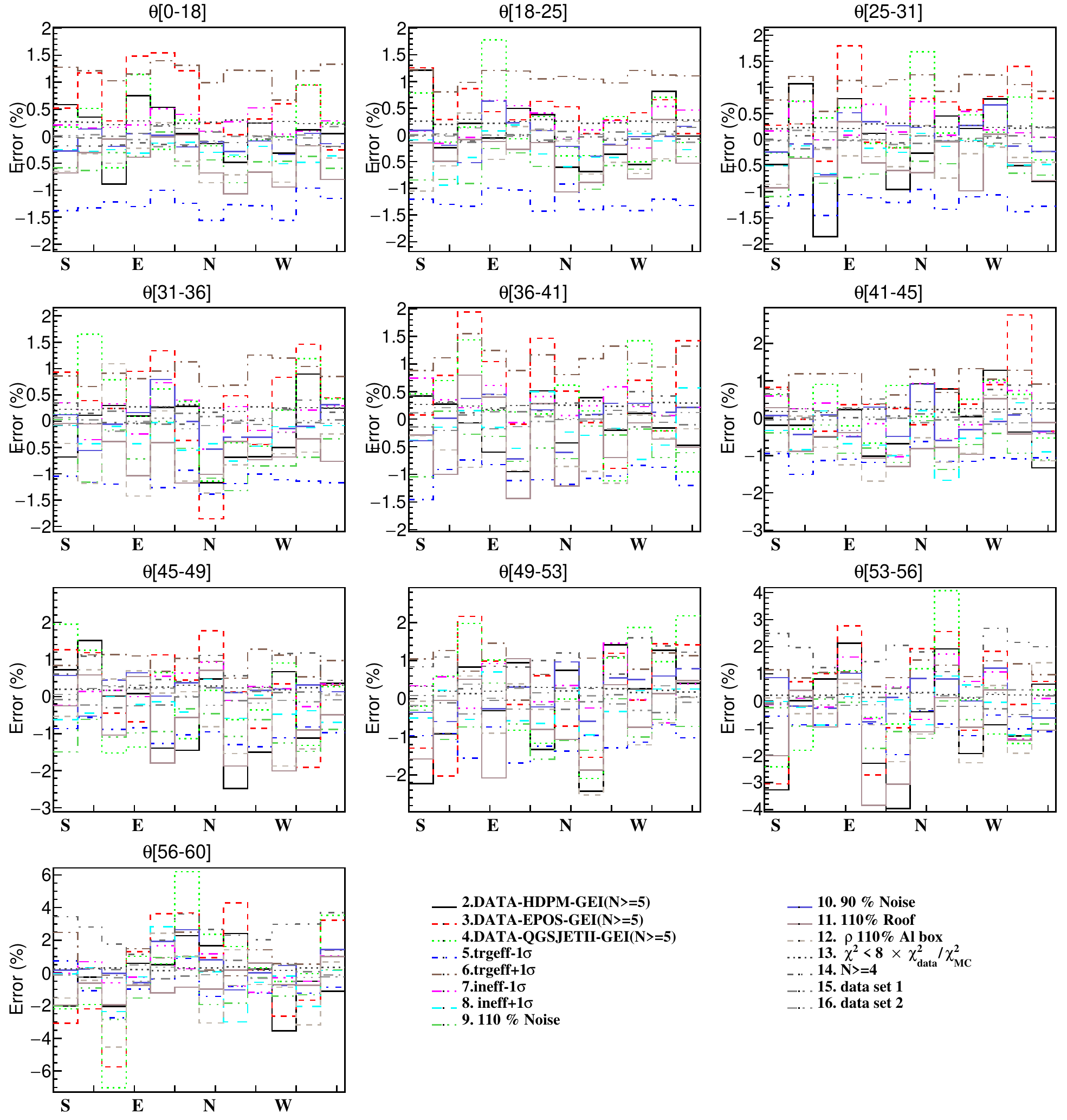} 
  \caption{Systematic errors in the estimation of muon flux for various ($\theta,\phi$) bins.} 
  \label{fig:systerr}
\end{figure}

The muon flux at different ($\theta,\phi$) bin can be affected by many systematics related to the uncertainties in the detector parameter and input muon spectrum to the GEANT4 input. To study these variations many input parameters are changed to propagate the uncertainties of those parameters to flux measurement and the muon flux is estimated. The various parameters modified in the MC to calculate the flux are,
$(i)$ To study the variation in the muon flux due to muon momentum spectrum, the muons spectrum generated by CORSIKA using various physics models (namely HDPM, EPOS-LHC and QGSJETII) is generated and propagated through the GEANT4.
$(ii)$ To account for the uncertainty in the estimated inefficiency, the binomial error in each 3\,cm\,$\times$\,3\,cm pixel is calculated and during the MC generation, the inefficiency is reduced and increased by 1$\sigma$.
$(iii)$ To account for the uncertainty in the estimated trigger efficiency, during the MC generation the trigger efficiency is decreased and increased by 1$\sigma$.
$(iv)$ The shower and electronic noise will affect the reconstruction efficiency of the muon. To check the variation of muon flux based on the estimated input noise to MC, the input noise is increased and decreased by 10\,$\%$, which is approximately the variation of noise during the whole period.
$(v)$ The complete geometry of the detector hall along with nearby buildings around the experimental hall has been included in the GEANT4 geometry, but there are uncertainties in the properties of the material of the wall and roof. To account these uncertainties, the roof thickness of the building is increased by 10\,$\%$, which modulates the muon spectrum at low energies.
$(vi)$ The low energy muons undergo multiple scattering in the materials in the RPC stack. To take care of that uncertainty, the density of the material inside the detector is changed by 10\,$\%$.
$(vii)$ The difference in the estimation of reduced-$\chi^{2}$ distribution in MC sample and data is also considered as a systematic, so the reconstructed muons from the MC is selected with scaled $\chi^{2}$/ndf.
$(viii)$ The muon flux is estimated by the reconstructed muons which are having minimum 5 layer hits, these criteria chosen to compensate large acceptance and larger track length. The flux is estimated for muons which are having minimum 4 layer hits and the difference is used as a systematic error on the selection criteria.
$(ix)$ The data sample is split into two sets as odd-numbered and even-numbered events, the flux is calculated for these two data samples separately.\\
The relative change in the muon flux for all these systematics is presented in figure \ref{fig:systerr}.

\section{Comparison of observed flux with CORSIKA and HONDA prediction}
\label{sec:compar}  

\begin{figure}
  \center
  \includegraphics[width=1.\linewidth]{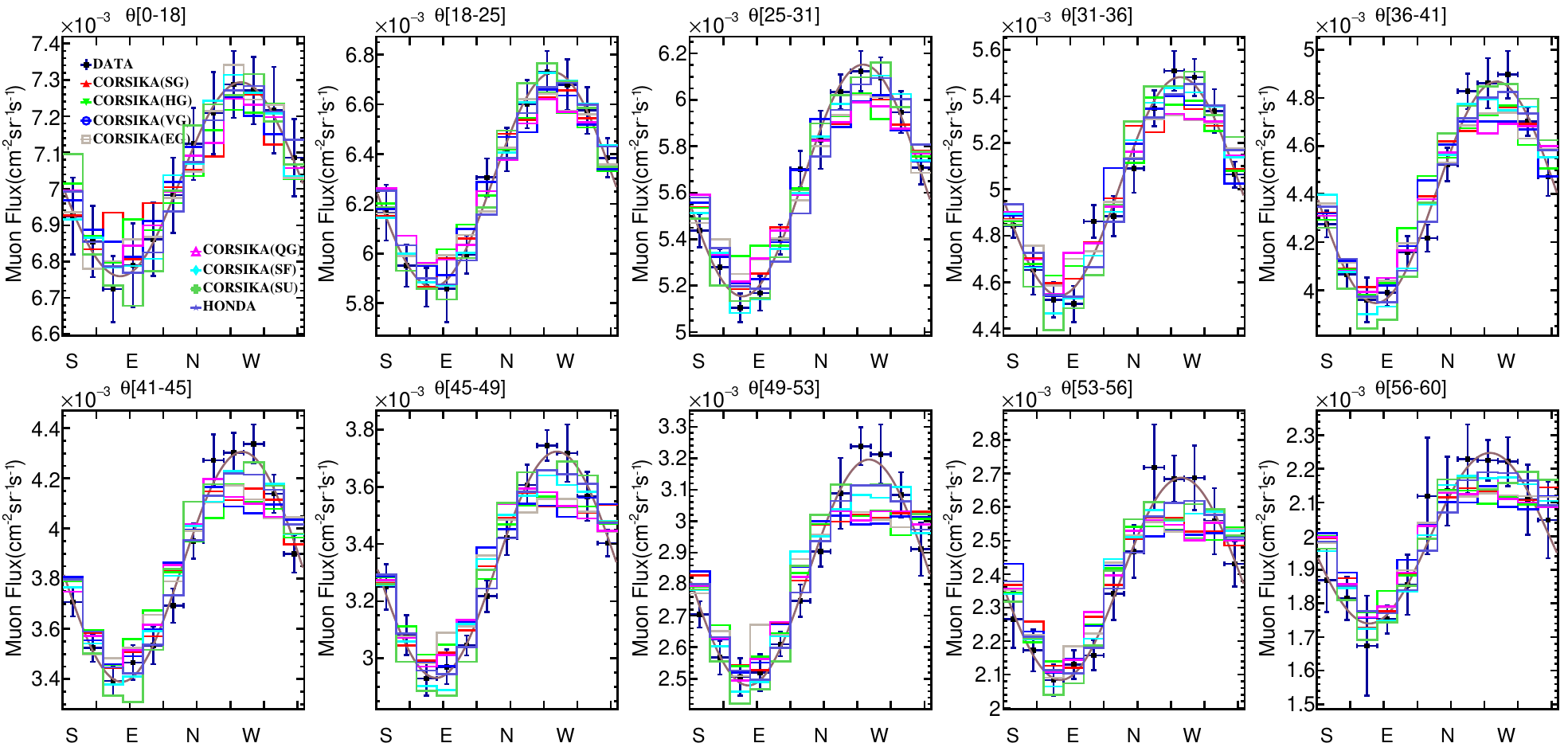} 
  \caption{Comparison of the azimuthal muon flux with CORSIKA and HONDA predictions.} 
  \label{fig:phytot}
\end{figure}

\begin{figure}
  \includegraphics[width=1.\linewidth]{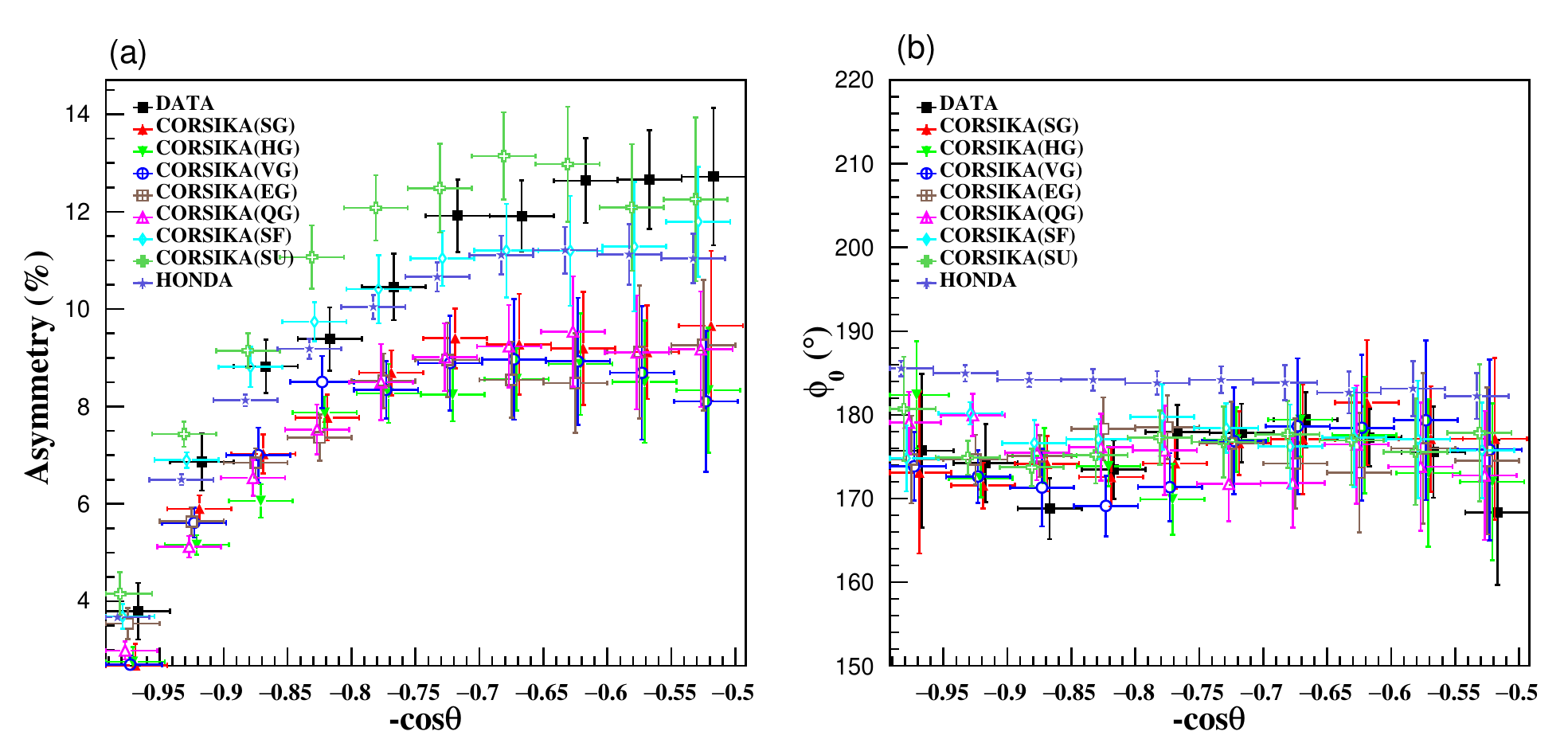} 
  \caption{(a) Asymmetry parameter for data, CORSIKA and HONDA, (b) $\phi_{0}$ parameter for data, CORSIKA and HONDA. (X-coordinates are shifted in different sets by different value for better visualisation)}
  \label{fig:par1_par2}
\end{figure}

The east-west asymmetry of cosmic ray muons at different locations on the earth is dependent on its geomagnetic latitude and longitude. A comparison of the results in the current study with the CORSIKA prediction for different physics interaction models and HONDA \cite{honda} prediction for the INO experimental site are discussed here. In order to compare the shape of the azimuthal dependent flux in the data with predictions, the CORSIKA and HONDA predictions are normalised by the observed flux. The observed flux in the data, CORSIKA with different input physics models at higher energies and same model at low energies (namely SIBYLL-GHEISHA (SG), VENUS-GHEISHA (VG) and HDPM-GHEISHA (HG)), EPOS-GHEISHA (EG) and QGSJETII-GHEISHA (QG)), with different input physics models at low energies and high energy model is fixed to one (namely SIBYLL-FLUKA (SF) and SIBYLL-urQMD (SU))\footnote{The shape of the azimuthal muon flux from CORSIKA using different high energy models are not differing much, so CORSIKA events are not generated for FLUKA and urQMD with different high energy models other than SIBYLL.} and HONDA predictions are shown in the figure \ref{fig:phytot} along with the following fit function,

\begin{equation}
  \label{eqn:equation4}
  f(\phi) = P_{0}(1 + Asin(-\phi + \phi_{0})),
\end{equation}

where parameter $A$ from the fit will give the amplitude of asymmetry of the fit function and $P_{0}$ is the average flux observed in the data. The fitted asymmetry parameters, $A$ and $\phi_0$ are shown in figure \ref{fig:par1_par2}(a) and \ref{fig:par1_par2}(b) respectively. The comparison of the measured result with the predictions are quantified using equation \ref{eqn:equation5},
\begin{equation}
  \label{eqn:equation5}
  \chi_{cos\,\theta}^{2} = \sum_{\phi_{i}=1}^{\phi_{i}=12} \frac{(I_{\phi_{i}}^{data} - I_{\phi_{i}}^{MC})^{2}}{\sigma_{\phi_{i}}^{2}},
\end{equation}

where $I_{\phi_{i}}^{data}$, $I_{\phi_{i}}^{MC}$ and $\sigma_{\phi_{i}}^{2}$ are the observed muon flux, the MC prediction of flux in various azimuthal bins and total error respectively. The calculated $\chi^{2}_{cos\theta}$ (for 12 data points) at cos\,$\theta$ from the equation \ref{eqn:equation5} are listed in table \ref{tab:table1}. It can be observed that the $\chi^{2}_{cos\theta}$ values for HONDA and CORSIKA with FLUKA low energy hadronic model has better agreement with the data in comparison with other input models of CORSIKA and not much variation with different high energy hadronic models.

\begin{table}[h]
  \begin{center}
    \begin{tabular}{|c|c|c|c|c|c|c|c|c|}
      \hline
      cos$\theta$ &  $\chi^{2}_{SG}$ & $\chi^{2}_{HG}$ & $\chi^{2}_{VG}$ & $\chi^{2}_{EG}$ & $\chi^{2}_{QG}$ & $\chi^{2}_{SF}$ & $\chi^{2}_{SU}$ & $\chi^{2}_{HONDA}$\\
      \hline
      1 -  0.95 & 9.14 &  5 &  4.54 &  3 &  2.36  &  1.72  &  5.2  &  1.78  \\
      0.95 -  0.9 & 4.67 &  9.91 &  6.8 &  9.12 &  10.1  &  4.91  &  4.69  &  6.96  \\
      0.9 -  0.85 & 14.3 &  33 &  15.2 &  19.1 &  23.5  &  5.75  &  6.62  &  19.9  \\
      0.85 -  0.8 & 17.3 &  16.5 &  14.4 &  27.6 &  24.7  &  10.8  &  29.3  &  17.6  \\
      0.8 -  0.75 & 21.6 &  34.8 &  31.1 &  22.2 &  22.2  &  19.4  &  26.1  &  11.5  \\
      0.75 -  0.7 & 22.2 &  31.7 &  36.9 &  29.6 &  27.7  &  13.5  &  22.6  &  13.5  \\
      0.7 -  0.65 & 36.8 &  28.6 &  46.7 &  34 &  28.8  &  22.4  &  21.1  &  6.91  \\
      0.65 -  0.6 & 39.4 &  38.1 &  48.5 &  39.8 &  38.6  &  25  &  26.4  &  13.8  \\
      0.6 -  0.55 & 19 &  23.5 &  22.5 &  19.9 &  21.6  &  13.8  &  12.1  &  7.4  \\
      0.55 -  0.5 & 9.5 &  14.6 &  13 &  9.34 &  8.07  &  3.61  &  6.55  &  5.17  \\
      \hline
    \end{tabular}
    \caption{The comparison of $\chi^{2}_{DATA-MC}$ (for 12 $\phi$ bins) for data with different MC predictions at different cos\,$\theta$ bins.}
    \label{tab:table1}
  \end{center}
\end{table}

The east-west asymmetry of cosmic muons increases with the zenith angle, $\theta$, as the geomagnetic rigidity for the west direction decreases with the increasing zenith angle while for the east, the rigidity increases. The measurements for asymmetry are compared with CORSIKA generated events with different low energy and high energy physics interaction models as well as the HONDA predictions at INO-site. The cosmic ray primaries with energy below 50\,GeV are mainly influenced by the geomagnetic field. In CORSIKA low energy interaction models are used up to 80\,GeV and beyond that energy, high energy interaction models are used. The changes in the high energy interaction models in the CORSIKA doesn't change the shape of the azimuthal spectrum. It was observed that the east-west asymmetry for different high energy models in CORSIKA doesn't differ much, the changes in the low energy interaction models predominantly affect the shape of the azimuthal flux. Among three low energy interaction models used in the study, the largest deviation in the east-west asymmetry from the data was observed with the GHEISHA interaction model. The simulation with other models, FLUKA and urQMD matches much better with observed data for most of the cos$\theta$ bins. The flux predictions from the HONDA are agreeing better with the observed asymmetry from the data in the higher cos$\theta$ bins. The phase ($\phi_{0}$) of the observed data is having a better match with all the models in CORSIKA generated events in comparison with HONDA.

\section{Conclusions}
\label{sec:conclu}
The azimuthal dependence of cosmic muon flux at different zenith angles has been studied using 2\,m $\times$\,2\,m RPC stack at IICHEP, Madurai. The east-west asymmetry of the atmospheric muons is well studied by many experiments at different locations on earth. Compared to them, the prototype RPC stack has been commissioned in a unique location, as it is very near to the geomagnetic equator. The cutoff rigidity at the geomagnetic equator is larger than any other place on earth. Due to this very high rigidity cutoff, the energy threshold for the primaries entering in the earth's atmosphere is much larger than other location. For example, an energy threshold of 38\,GeV (12\,GeV) is experienced by the primaries entering the atmosphere at 60$^{\circ}$ zenith angle and are coming from the east (west) direction. Due to this huge difference in the energy threshold, the east-west asymmetry near the geomagnetic equator is very large. The present results of east-west asymmetry is most comparable with CORSIKA simulation, where FLUKA and urQMD are used as low energy hadronic interaction models. The CORSIKA simulation with GHEISHA as a low energy input model shows less east-west asymmetry in comparison with data. The azimutal distribution of muons does not depend much on the chosen high energy interaction model assumed in the simulations. The HONDA predictions are comparable with data in the higher cos$\theta$ bins. The systematic study of the muon flux has been estimated by considering the uncertainties in the detector parameters and various physics models. This result Will contribute to the reduction of the systematics in the neutrino flux due to pion and kaon productions in the primary interactions in the atmosphere. The detector parameters like inefficiencies, multiplicities, time and position resolutions observed in the data are going to be an input to the INO-ICAL simulation code in the digitization stage to make it more realistic.
\begin{acknowledgements}
  The authors would like to thank Dr.P.K. Mohanty and Mr. Hariharan from GRAPES-3 experiment for providing angle dependent primary cutoff rigidity values for the experimental site. The authors also acknowledge crucial contributions by A. Bhatt,  S.D. Kalmani,  S. Mondal,  P. Nagaraj, Pathaleswar,  M.N.  Saraf,  R.R.  Shinde,  Dipankar  Sil, S.S. Upadhya, P. Verma, E. Yuvaraj, S.R. Joshi, Darshana Koli, S. Chavan, N. Sivaramakrishnan, B. Rajeswaran, Rajkumar Bharathi in setting up the detector, electronics and the DAQ systems.
\end{acknowledgements}

\end{document}